%% file: main.tex
\documentclass[conference]{IEEEtran}
\IEEEoverridecommandlockouts
\usepackage{cite}
\usepackage{amsmath,amssymb,amsfonts}
\usepackage{algorithmic}
\usepackage{graphicx}
\usepackage{textcomp}
\usepackage{xcolor}
\usepackage{tikz}
\usepackage{booktabs}
\usepackage{siunitx}
\usepackage{subcaption}
\usepackage{listings}
\usepackage{color}
\usepackage{multirow}
\usepackage{afterpage}
\usepackage[framemethod=TikZ]{mdframed}
\usepackage{enumitem}
\usepackage{url}
\usepackage{microtype}
\usepackage{balance}

\newcommand*\circled[1]{
    \tikz[baseline=(char.base)]{
        \node[shape=circle, fill=black, text=white, inner sep=0.1pt, minimum size=1em, text centered] (char) {\footnotesize #1};
    }
}

\definecolor{templateColor}{rgb}{0.2,0.6,0.2}
\newcommand{\code}[1]{\lstinline[basicstyle=\ttfamily]{#1}}

\newcommand{\placeholder}[1]{#1}

\definecolor{codegreen}{rgb}{0,0.4,0}       
\definecolor{codegray}{rgb}{0.3,0.3,0.3}   
\definecolor{codepurple}{rgb}{0.2,0,0.4}   
\definecolor{backcolour}{rgb}{0.98,0.98,0.95} 
\definecolor{codemagenta}{rgb}{0.55,0,0.55} 

\lstdefinestyle{mystyle}{
  backgroundcolor=\color{backcolour},   
  commentstyle=\color{codegreen},
  keywordstyle=\color{codemagenta},
  numberstyle=\tiny\color{codegray},
  stringstyle=\color{codepurple},
  basicstyle=\ttfamily\scriptsize,
  breakatwhitespace=false,         
  breaklines=true,                 
  captionpos=b,                    
  keepspaces=true,                 
  numbers=left,                    
  numbersep=5pt,                  
  showspaces=false,                
  showstringspaces=false,
  showtabs=false,                  
  tabsize=2,
  frameshape={RYR}{Y}{Y}{RYR},
  xrightmargin=0.5em,
  }

\lstdefinelanguage{bnf}{
  morekeywords={askit_api,ask,define,prompt_template,param_types,examples,example,input},
  morecomment=[l]{//},
  morecomment=[s]{/*}{*/},
  morestring=[b]',
  morestring=*[d]",
  moredelim=[s][\color{blue}]{\{\{}{\}\}},  
}

\lstdefinelanguage{TypeScript}{
  morekeywords={
    ask,
    define,
    abstract,
    any,
    type,
    string,
    number,
    boolean, 
    const,
    let,
    await,
    void},
  morecomment=[l]{//},
  morecomment=[s]{/*}{*/},
  morestring=*[d]",
  morestring=*[d]',
  moredelim=[s][\color{blue}]{\{\{}{\}\}}
}

\lstdefinelanguage{python}{
  morekeywords={
    dict,
    List,
    define,
    int,
    str,
    compile
  },
  morecomment=[l]{\#},
  morecomment=[s]{'''}{'''},
  morecomment=[s]{"""}{"""},
  morestring=*[d]",
  morestring=*[d]',
  moredelim=[s][\color{blue}]{\{\{}{\}\}},
}

\def\BibTeX{{\rm B\kern-.05em{\sc i\kern-.025em b}\kern-.08em
    T\kern-.1667em\lower.7ex\hbox{E}\kern-.125emX}}

\begin{document}

\title{\emph{AskIt}: Unified Programming Interface for Programming with Large Language Models}         

\author{\IEEEauthorblockN{Katsumi Okuda}
\IEEEauthorblockA{\textit{CSAIL, MIT}\\
Cambridge, USA \\
okuda@csail.mit.edu}
\IEEEauthorblockA{\textit{Mitsubishi Electric Corporation}\\
Amagasaki, Japan }
\and
\IEEEauthorblockN{Saman Amarasinghe}
\IEEEauthorblockA{\textit{CSAIL, MIT}\\
Cambridge, USA \\
saman@csail.mit.edu}
}

\maketitle

\begin{abstract}
Large Language Models (LLMs) exhibit a unique phenomenon known as \emph{emergent abilities}, demonstrating adeptness across numerous tasks, from text summarization to code generation.
While these abilities open up novel avenues in software design and crafting, their incorporation presents substantial challenges.
Developers face decisions regarding the use of LLMs for directly performing tasks within applications as well as for generating and executing code to accomplish these tasks.
Moreover, effective prompt design becomes a critical concern, given the necessity of extracting data from natural language outputs.
To address these complexities, this paper introduces \emph{AskIt}, a domain-specific language (DSL) specifically designed for LLMs.
AskIt simplifies LLM integration by providing a unified interface that not only allows for direct task execution using LLMs but also supports the entire cycle of code generation and execution. 
This dual capability is achieved through (1)~type-guided output control, (2)~template-based function definitions, and (3)~prompt generation for both usage modes.
Our evaluations underscore AskIt's effectiveness.
Across 50 tasks, AskIt generated concise prompts, achieving a 16.14~\% reduction in prompt length compared to benchmarks.
Additionally, by enabling a seamless transition between using LLMs directly in applications and for generating code, AskIt achieved significant efficiency improvements, as observed in our GSM8K benchmark experiments.
The implementations of AskIt in TypeScript and Python are available at \url{https://github.com/katsumiok/ts-askit} and \url{https://github.com/katsumiok/pyaskit}, respectively.
\end{abstract}

\begin{IEEEkeywords}
domain specific language, code generation, large language model, software engineering, artificial intelligence
\end{IEEEkeywords}
  
\section{Introduction}

Recent studies \cite{wei2022emergent} have unveiled the remarkable abilities of Large Language Models (LLMs), which become increasingly pronounced with model scaling. These abilities span a wide range of tasks, including arithmetic operations~\cite{saxton2019analysing}, question answering~\cite{Lewis_2019}, text summarization~\cite{Narayan_2018}, language translation~\cite{Brown_2020}, paraphrasing \cite{prakash2016neural}, text prediction~\cite{Srivastava_2018}, and code generation~\cite{chen2021evaluating,li2022competition,fried2023incoder,li2023starcoder,2023code}.
Intriguingly, these capabilities are not imparted explicitly but are organically cultivated through vast exposure to natural language data during training.
This phenomenon, termed \emph{emergent abilities}, distinguishes LLMs. The notion of \emph{emergent abilities} is captivating, hinting that with further advancements in language models, even more sophisticated capabilities may emerge.

The rise of these emergent abilities holds significant implications for software development, potentially altering the very methods by which software is crafted. Developers can incorporate LLMs within applications to handle tasks such as question answering, text summarization, or language translation. Another application of LLMs is in code generation. Tools like Jigsaw~\cite{10.1145/3510003.3510203} and Codex/Copilot~\cite{chen2021evaluating} harness LLMs to convert natural language descriptions into code. Even without these specific tools, developers can leverage LLM-based chatbots, like ChatGPT based on GPT-4~\cite{openai2023gpt4}, BingAI, and Bard, for the same purpose.

However, integrating LLMs into software development is not without challenges. One primary decision developers face is whether to embed the LLM directly into the application or employ it for code generation. The distinction between these two applications is stark, making it laborious to transition between them later.
For instance, while one could incorporate an LLM directly into an application to sort a list of numbers, another approach would be to utilize the LLM to generate the code for sorting. Choosing between these methodologies post-decision can be laborious.
These methodologies differ significantly, and altering the chosen approach subsequently demands considerable effort.

Moreover, regardless of the approach, developers must devise effective prompts, extract pertinent data from the LLM's output, and then process it. If the application integrates an LLM for its functionality, code must be written to parse the LLM's response — a non-trivial task given the natural language format. Hence, specifying the desired data format within the prompt is often adopted to ease response parsing. Yet, this necessitates precise, task-specific prompt design. When LLMs are used for code generation, the resultant code must be manually integrated into the application.

In response to these challenges, we present \emph{AskIt}: a domain-specific language (DSL) tailored for LLMs. AskIt offers a harmonized programming interface across varied tasks, featuring (1) type-guided output control, (2) template-based function definitions, (3) code generation capabilities.
The type-guided output control obviates the need for data format specification within natural language prompts, eliminating the intricate prompt engineering previously essential for response extraction. Template-based function definitions allow developers to craft functions leveraging an LLM, using prompts tailored to specific tasks. Such templates can accept input parameters that seamlessly map to the defined function's parameters. With code generation, there's no demarcation between integrating an LLM into an application and using it for code generation, allowing effortless transitions between the two without adjusting the prompt template.

We demonstrate AskIt's applicability across a wide range of LLM tasks.
By using AskIt to implement 50 common tasks, we show that AskIt can generate 7.56 lines of TypeScript code and 6.52 lines of Python code on average.
We also confirmed that AskIt can reduce the length of prompt by 16.14~\% on average compared to the original prompts used in the OpenAI Evals~\footnote{\url{https://github.com/openai/evals}} benchmark.

Additionally, we measured the speedup of functions defined with AskIt when we transitioned from using an LLM as part of the application to executing equivalent functions generated by the LLM. An experiment with the GSM8K benchmark~\cite{cobbe2021training} revealed that generated functions by AskIt with GPT-4 achieved a speedup of 275,092.55x in TypeScript and 6,969,904.73x in Python, respectively, compared to the same functions using GPT-4 as part of the application.

The contributions of this paper are summarized as follows:

\begin{enumerate}
\item We categorize tasks in terms of whether they are directly answerable by LLMs and whether they are codable by LLMs and identify the challenges faced when integrating LLMs into software development. 
\item We introduce a unified programming interface tailored for LLMs to accommodate for codable tasks, directly answerable tasks, and tasks that fall into both categories.
This interface features \emph{type-guided output control}, \emph{template-based function definitions}, and \emph{code generation} capabilities.
This eliminates the boundary between the direct application integration of an LLM and its use for code generation.
\item We implemented AskIt in both a statically typed language (TypeScript) and a dynamically typed language (Python) and evaluated them across a diverse set of tasks, showcasing its potency in code generation and efficiency in prompt reduction.
\end{enumerate}


\section{Motivating Examples} \label{sec:motivating_examples}

To underscore the need for a unified and streamlined approach to incorporating LLMs into programming tasks, this section explores two distinct applications that could benefit from LLMs.
The first demonstrates the potential of LLMs in sentiment analysis of product reviews, while the second discusses a file access task that stores the results of the sentiment analysis in a local file system.
In both cases, the software developer must craft a prompt and either interpret the response from the LLM or integrate the generated code into their source code.

\subsection{Examples}
\subsubsection{Using an LLM as Part of an Application}

Consider a scenario in which a developer is writing a program to analyze the sentiment of product reviews. Although this task traditionally relies on complex natural language processing pipelines or machine learning models, an LLM like GPT-4 can significantly simplify the process. With an appropriately crafted prompt, the LLM can interpret and deduce the sentiment behind a given review.

Below is a simplified pseudo-code representation:

\noindent
\begin{minipage}\linewidth
\begin{lstlisting}[language=python, style=mystyle]
review = "The product is fantastic. It exceeds all my expectations."
prompt = "Determine the sentiment of this review: '" + review + "'. The final sentiment should be enclosed in [ and ] like [negative]."
response = LLM.predict(prompt)  # response: "The sentiment of the review is [positive]."
sentiment = parse_sentiment(response)  # sentiment: "positive"
\end{lstlisting}
\end{minipage}
where \code{\#} denotes a comment.
Line 1 initializes the review. In practice, this would typically be sourced from a database or another data source, but it's hardcoded here for illustrative purposes. 
Line 2 crafts the prompt by integrating the review with a templated structure.
Line 3 engages an LLM, processing the prompt to generate a response.
Finally, Line 4 extracts the sentiment from the response.

This scenario introduces two major challenges:
\begin{itemize}
\item \textbf{Parsing the LLM's response:} Developers must write code to extract the sentiment from the LLM's natural language output. Due to potential variability in the LLM's responses, based on the prompt and its inherent behavior, this extraction is far from trivial. 
\item \textbf{Crafting the prompt:} This task requires a deep understanding of natural language processing and familiarity with potential LLM responses. By specifying the desired response format with [ and ], the developer can make the subsequent parsing easy.
\end{itemize}
While the provided example is straightforward, with relatively simple prompt construction and response parsing, more complex problems can introduce challenges. In these scenarios, techniques like Chain of Thought (CoT) \cite{NEURIPS2022_8bb0d291} become essential, guiding the LLM to produce better responses. If the expected output involves multiple facets, such as a list or several values, crafting the prompt and response parsing can become more complex, demanding additional effort from the developer.
Thus, while LLMs like GPT-4 offer powerful capabilities, developers must skillfully craft their prompts and parsing methods to ensure accuracy and reliability.

\subsubsection{Using an LLM to Generate Code}

Expanding upon the sentiment analysis task, let's delve into a scenario where the results of the sentiment analysis need to be saved to a local CSV (comma-separated values) file. Although this is not an inherent function of LLMs, their ability to generate relevant code exemplifies their adaptability.

Typically, LLMs generate code snippets in response to a high-level description provided by the user. For instance, a developer might employ an LLM, seeking to generate Python code that saves sentiment analysis results into a local file. This interaction often takes place within platforms like ChatGPT, Bard, or BingAI. Here, the developer would pose a task description, such as "Generate a Python function to log a product review and its associated sentiment into a specific CSV file."

\begin{figure}[t]
\centering
\includegraphics[width=0.48\textwidth]{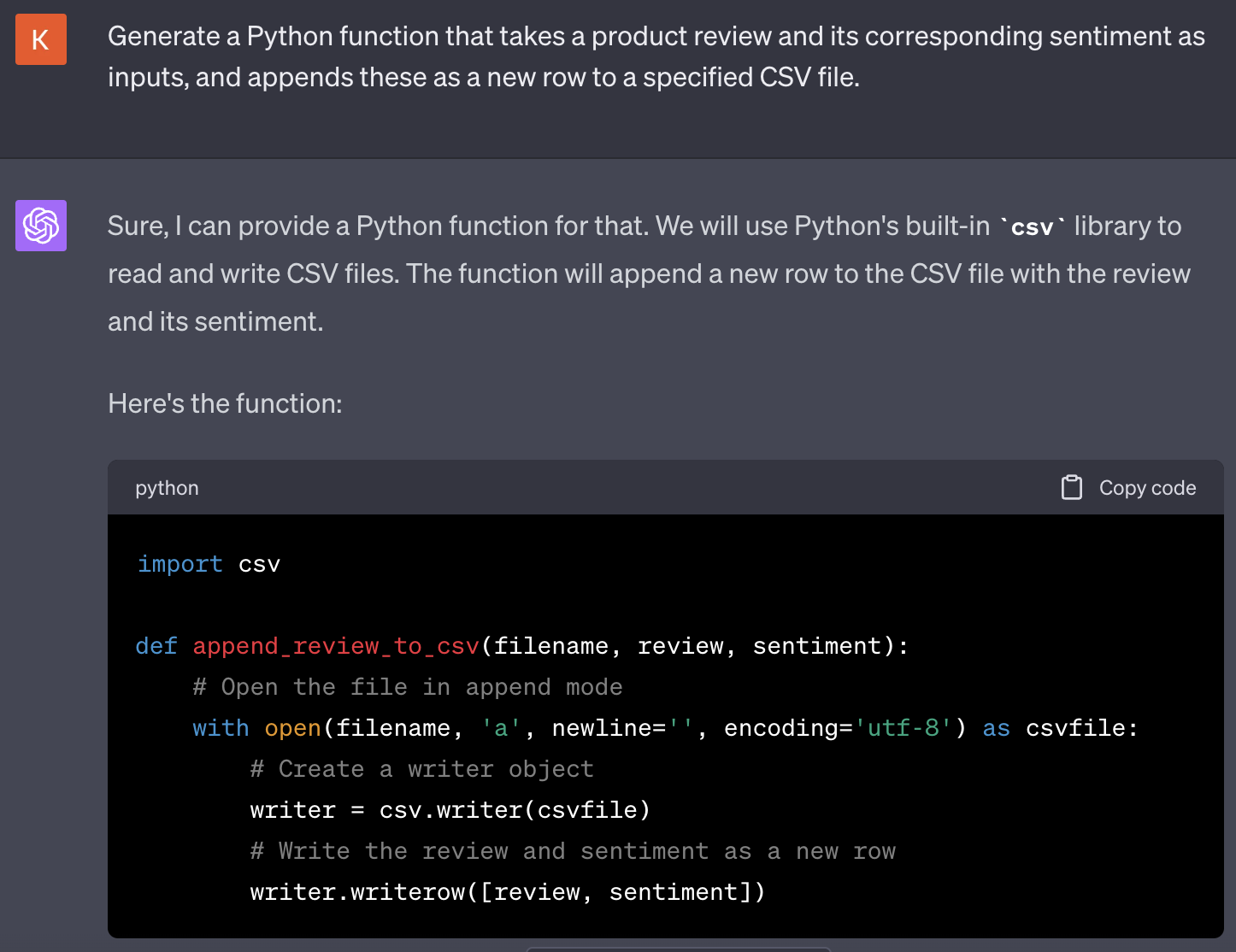} 
\caption{Code generation by LLM (ChatGPT)} \label{fig:chatgpt}
\end{figure}

Figure~\ref{fig:chatgpt} illustrates an interaction with ChatGPT. After the developer inputs the task description, ChatGPT responds with a pertinent code snippet. The developer can then manually copy and incorporate this snippet into their existing codebase. This generated function opens the specified CSV file in append mode and saves the 'review' and 'sentiment' as a new row. However, it's important to note that ChatGPT cannot execute this generated code directly since it doesn't have access to the local file system. Thus, developers need to manually integrate this snippet into their software environment. Sometimes, this incorporation requires tweaks to make the newly added code align with the existing codebase. While this code generation approach is useful, it still demands manual intervention which could be made more efficient. 

\begin{figure}[t]
\centering
\includegraphics[width=0.4\textwidth]{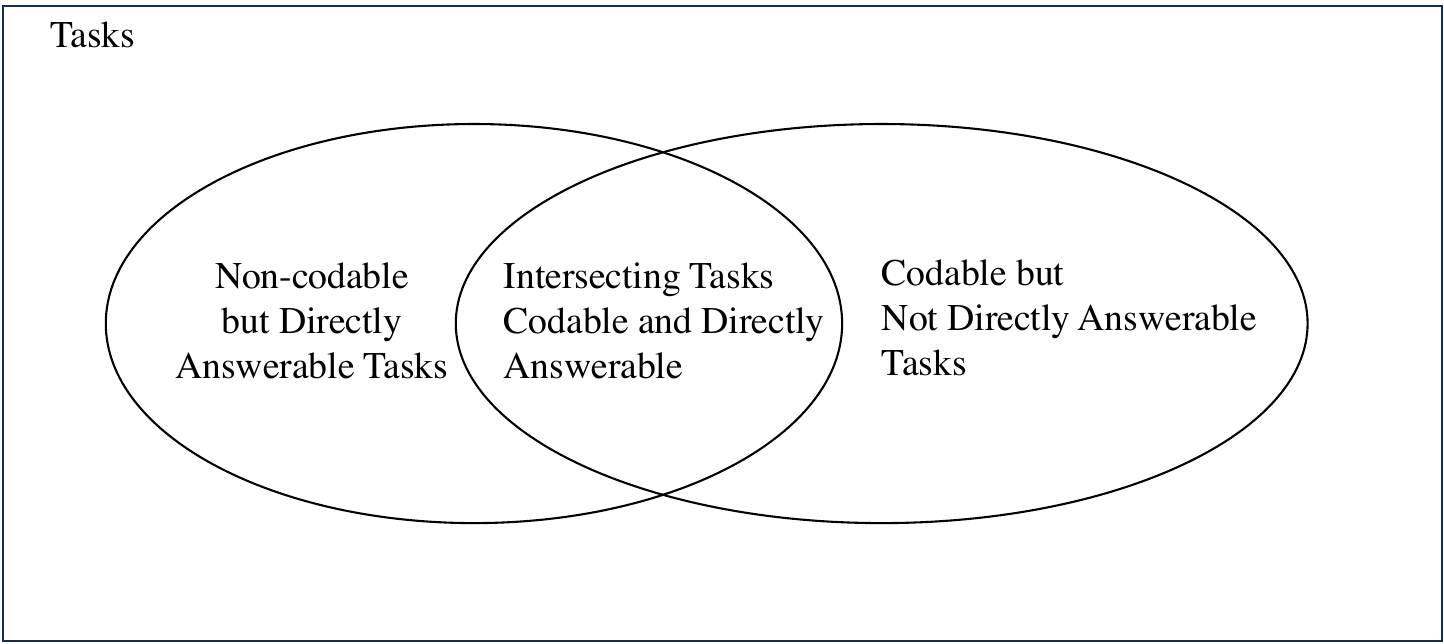} 
\caption{Classification of Tasks} \label{fig:venn}
\end{figure}

\subsection{Classification and Examples of Problem Types for LLMs}

In the previous examples, we presented two distinct types of tasks that can be addressed or facilitated by LLMs.
One involves using LLMs as part of an application, while the other entails using LLMs to generate code.
To optimize the use of LLMs, we categorize tasks based on the following two dimensions:
\begin{itemize}
\item \textbf{Directly Answerable or Not:} Determines whether the LLMs can directly answer the task.
\item \textbf{Codable or Non-Codable:} Indicates whether LLMs can generate code to do the task.
\end{itemize}
These dimensions are orthogonal to each other.
We can leverage LLMs if either of the dimensions holds true.
In other words, we can integrate LLMs into an application if the task is directly answerable, and we can employ LLMs to implement code if the task is codable.
Given these dimensions, we can group tasks into three categories: non-codable but directly answerable tasks, intersecting tasks, and codable but not directly answerable tasks, as illustrated in Figure~\ref{fig:venn}.
The prior two examples fall under the categories of non-codable but directly answerable tasks and codable but not directly answerable tasks, respectively.
The sentiment analysis task represents a non-codable but directly answerable task because LLMs can immediately address it, whereas traditional programming methods might not achieve comparable accuracy.
Conversely, the file access task is codable but not directly answerable because traditional programming techniques can handle it, but it isn't straightforwardly answered by LLMs.

It's worth noting that some tasks can be addressed both directly by LLMs or through LLM-generated code.
For such tasks, either solution may be chosen.
Take, for example, the mathematical query "What is 7 times 8?". This could be resolved either by employing LLMs within an application or by using LLMs to generate code.
Generally, intersecting tasks exhibit superior performance when tackled by generated code than when directly addressed by LLMs.
However, the delineation of tasks isn't always evident and can be ambiguous.
The boundaries separating the three categories are often blurred.
As we will illustrate with our experimental results, certain mathematical problems are answerable by LLMs but resist coding by LLMs.

The challenge arises from the distinct implementation needs of the two approaches.
When LLMs are incorporated into an application, we must craft code to parse the LLM responses.
Conversely, when LLMs are used for code generation, the resulting code must be manually integrated into our source code.
A unified interface for these strategies is absent, complicating the transition between the two methods.
Should such a unified interface exist, transitioning between the two techniques would be more straightforward.
This adaptability is essential, especially given the inherent ambiguity in task classification.
Furthermore, as LLMs continue to evolve, the borders defining the three categories are bound to shift.

\section{Design and Implementation} \label{sec:askit}

\subsection{Overview}

Our Domain Specific Language (DSL), \emph{AskIt}, offers two APIs: \code{ask} and \code{define}. They serve as a unified interface by borrowing the syntax of function calls from the host programming language. Hence, they can be used wherever function calls are permitted. These APIs address a wide array of tasks, such as non-codable yet directly answerable tasks, intersecting tasks, and codable but not directly answerable tasks, as detailed in the previous section. The features of the APIs are as follows: (1) Type-guided output control, (2) template-based function definitions, and (3) code generation from the unified interface.

As a proof of concept, we implemented TypeScript and Python versions of AskIt. The DSL compiler is fashioned as a TypeScript compiler plugin for TypeScript and as a Python library for Python. AskIt compiler and runtime synthesize the prompt for the LLMs and the parser for the response based on the type information of the function and variables embedded in the template expression. It also generates the function that implements codable tasks.
In the following, we illustrate these features using the same examples provided in the previous section. Although we use TypeScript for the examples, a similar syntax can be adopted for Python. The implementation in Python is discussed later in \ref{sec:dynamic}.

\subsubsection*{Type-Guided Output Control}

A typical example of a non-codable yet directly answerable task is determining the sentiment of a review. We assume the sentiment can be either positive or negative. Instead of detailing the expected output format in the prompt, we can specify the expected output type in the DSL. For instance, the following code is valid AskIt code for the sentiment analysis task:

\begin{lstlisting}[language=TypeScript, style=mystyle]
let sentiment = await ask<'positive' | 'negative'>('What is the sentiment of the following review: "The product is fantastic. It exceeds all my expectations."');
\end{lstlisting}
\noindent
Here, \code{ask} is an API that accepts a prompt and returns a response. The response's type is indicated in the type parameter of \code{ask}. In this instance, \code{'positive'|'negative'} is a union type, which consists of two string literal types and signifies that the response is either \code{'positive'} or \code{'negative'}, 
This type information aids in generating the prompt for the LLMs. After executing the code, the variable \code{sentiment} will be assigned the value \code{'positive'}.
\code{await} is a keyword that indicates the asynchronous execution of the \code{ask} API. The \code{ask} API returns a promise, and the \code{await} keyword is used to wait for the promise to be resolved.

Moreover, a prompt can be parameterized by using a prompt template as an argument for \code{ask}. Using a prompt template, the example above can be rewritten as:

\begin{lstlisting}[language=TypeScript, style=mystyle]
let sentiment = await ask<'positive' | 'negative'>('What is the sentiment of {{review}}?');
\end{lstlisting}
\noindent
Here, \code{review} is a string type variable. \code{\{\{} and \code{\}\}} mark the start and end of a variable in the prompt template, respectively. The variable \code{review} captures the symbol declared in the same scope.

\subsubsection*{Template-based Function Definitions}

In practical software development, the same task often needs replication. AskIt introduces a mechanism to formulate a function to repeatedly perform the same task. For instance, a function can be designed to return the sentiment of a review:
\vspace{2mm}
\begin{lstlisting}[language=TypeScript, style=mystyle]
let getSentiment = define<'positive' | 'negative'>('What is the sentiment of {{review}}?');
\end{lstlisting}
\noindent
Here, \code{define} is an API that accepts a prompt template and returns a function. The type parameter of \code{define} determines the function's return value. The function's parameter is defined in the prompt template. In this example, the function receives a variable named \code{review}.
The parameter in the template prompt corresponds to the parameter of the function defined with the same name.
By giving an actual argument, the defined function can be called as follows:
\begin{lstlisting}[language=TypeScript, style=mystyle]
let sentiment = await getSentiment({review: 'The product is fantastic. It exceeds all my expectations.'});
\end{lstlisting}
\noindent
Upon execution, \code{sentiment} will hold the value \code{'positive'}.

\subsubsection*{Code Generation}

As discussed previously, LLMs can be employed for code generation. For example, LLMs can be used to implement a function that appends a review and its sentiment to a CSV file. There's no need to use different APIs for code generation. Our cohesive interface enables function generation:

\begin{lstlisting}[language=TypeScript, style=mystyle]
let appendReviewToCsv = define<void>('Append {{review}} and {{sentiment}} as a new row in the CSV file named {{filename}}');
\end{lstlisting}

\noindent
Here, \code{filename}, \code{review}, and \code{sentiment} are variables. The above code can be invoked anywhere in the source code:

\noindent
\begin{minipage}\linewidth
\begin{lstlisting}[language=TypeScript, style=mystyle]
appendReviewToCsv({
    filename: 'reviews.csv',
    review: 'The product is fantastic. It exceeds all my expectations.',
    sentiment: 'positive'
});
\end{lstlisting}
\end{minipage}
\noindent
Before the code's execution, the DSL compiler, with the assistance of an LLM, will generate a function that appends a review and its sentiment to a CSV file.

\subsection{Syntax}

The AskIt syntax builds upon the function call structure of the host programming language. Listing~\ref{lst:syntax} presents the AskIt syntax tailored for TypeScript.
For this grammar, we assume that \code{TYPE}, \mbox{\code{STRING_LITERAL},} \code{IDENTIFIER}, and \code{CONSTANT_EXPRESSION} are non-terminal symbols.
They denote the type, string literal, identifier, and constant expression in the host programming language, respectively.

\begin{lstlisting}[style=mystyle, language=bnf, caption={Syntax of AskIt for TypeScript}, label={lst:syntax}, float=t, xleftmargin=2mm]
askit_api ::= ask | define
ask ::= "ask" "<" TYPE ">" "("  prompt_template examples? ")"
define ::= "define" "<" TYPE param_types? ">" "(" prompt_template examples? examples? ")"
prompt_template ::= STRING_LITERAL
param_types ::= "," "{" IDENTIFIER ":" TYPE ("," IDENTIFIER ":" TYPE)* "}"
examples ::= "," "[" example ("," example)* "]"
example ::= "{" "input" ":" input "," "output" ":" CONSTANT_EXPRESSION "}"
input ::= "{" IDENTIFIER ":" CONSTANT_EXPRESSION ("," IDENTIFIER ":" CONSTANT_EXPRESSION)* "}"
\end{lstlisting}

The primary APIs provided by AskIt are \code{ask} and \code{define} (Line~1). 
The \code{ask} API takes the response type of an LLM as a type parameter and takes a prompt template as a function parameter. Optionally, it can also take examples of the task's input-output pairs (Line~2). These examples facilitate few-shot learning~\cite{Brown_2020} and provide a way of Programming by Example (PBE)~\cite{10.1145/1925844.1926423,shaw1975inferring}.
In contrast, the \code{define} API takes the LLM's response type and optional parameter types as type parameters (Line~3). Like \code{ask}, \code{define} can incorporate a prompt template and examples. Moreover, \code{define} can accept two sets of input-output examples. While the first set is used for few-shot learning, the second set is utilized for validating the generated code.

The prompt template is essentially a string literal (Line~4), but it can have placeholders for variables.
These placeholders are identifiers enclosed between \code{\{\{} and \code{\}\}}.
The variable name within this placeholder should be a valid identifier of the host programming language.

Parameter types are key-value pairs listed within \code{\{} and \code{\}}, separated by commas (Line~5). Here, the key signifies the variable name, and the value represents its type in the host language.

Examples consist of input-output pairs. They are enclosed within \code{[} and \code{]} and separated by commas (Line~6). Each example, bounded by \code{\{} and \code{\}}, has an \code{input} key, which links to a task input, and an \code{output} key, pointing to the task output (Line~7). An input is a collection of key-value pairs, where the key is a variable name and the value is a constant expression defined by the host language. The output is a standalone constant expression.

\subsection{Computation Flow}

\begin{figure}[t]
  \centering
  \includegraphics[width=0.48\textwidth]{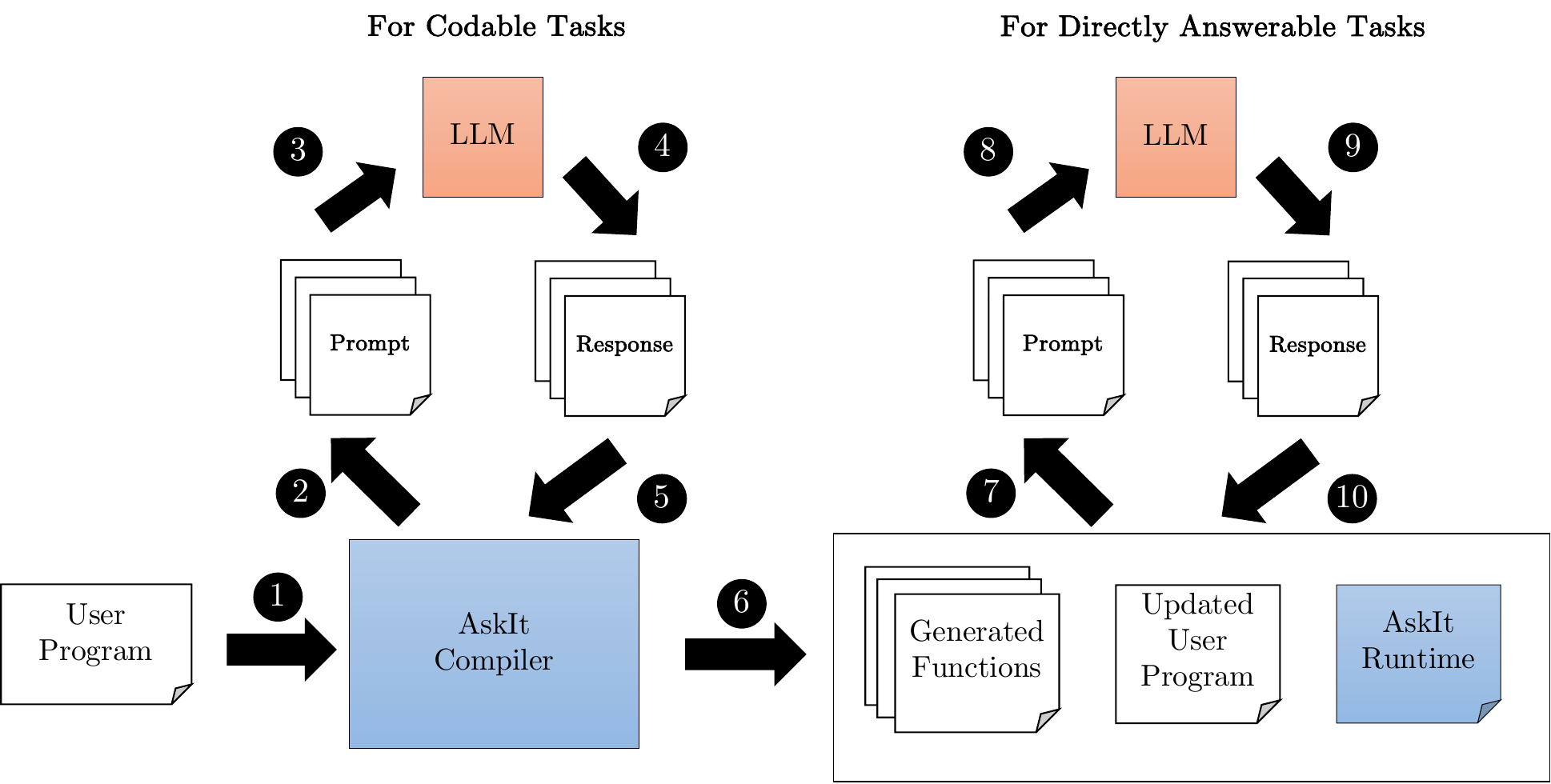}
  \caption{Computational flow of \emph{AskIt} DSL for statically typed language} \label{fig:flow}
  \end{figure}
  
In a proof of concept, we implemented AskIt for TypeScript and a DSL compiler as a TypeScript compiler plugin.
The DSL compiler is triggered when the TypeScript compiler compiles the source code.
The source code is written in TypeScript extended with our DSL, and the output of the DSL compiler is TypeScript code.

The computational flow of the DSL compiler is illustrated in Figure~\ref{fig:flow}.
The left side of the figure shows the computational flow at the compilation time, and the right side shows the computational flow at the runtime.
When the DSL compiler is triggered, it traverses the Abstract Syntax Tree (AST) of the source code and converts the AskIt APIs to specific functions written in TypeScript (\circled{1}).
If the call to \code{define} is detected and it is a codable task, the DSL compiler generates a function (\circled{2}--\circled{6}). First, the DSL compiler generates a prompt to ask an LLM to code the task (\circled{2}). Then, the prompt is passed to the LLM (\circled{3}). The LLM generates a response with a code for the task (\circled{4}), and the DSL compiler receives the response and parses the response to extract the generated code (\circled{5}). Finally, the DSL compiler validates the code and stores it. At the same time, a call to \code{define} is replaced with a reference to the generated function (\circled{6}).

The DSL compiler also updates calls to \code{ask} and \code{define} even if they are not codable tasks.
The DSL compiler extracts the type information from the type parameter of \code{ask} and \code{define} and encodes them into data to be used to generate a prompt at runtime (\circled{6}).

The user program at the runtime consists of the generated functions, the updated user program and the DSL runtime.
When the user program is executed, the updated user program calls the generated functions if the task is codable.
If the call to \code{ask} or functions defined by \code{define} is detected, the DSL runtime generates a prompt based on the type information extracted at the compilation time (\circled{7}). Then, the prompt is passed to the LLM (\circled{8}). The LLM generates a response that contains the data in the specified type (\circled{9}), and the DSL runtime receives the response and parses it to extract the answer(\circled{10}).

\subsection{Code Generation for Codable Tasks}\label{sec:code_generation}

Our DSL compiler generates a function that implements the task specified by the prompt template passed to the \code{define} call.
All calls to \code{define} are examined to determine if the task is codable. If it is, the DSL compiler generates a function and replaces the call to \code{define} with the generated function. As a result, calls to generated functions are executed without invoking the LLM at runtime.

We provide two ways to specify the task for codability by LLMs. The first method allows users to specify the name of a source file containing the call to \code{define}. In this case, the DSL compiler generates functions for all the calls to \code{define} in the specified source file. The second, more granular method lets the user specify the name of the function to be generated. This function name corresponds to the variable name to which the result of the \code{define} call is assigned.

For all calls to \code{define} designated as codable, the DSL compiler follows these steps to generate a function with an LLM:

\begin{enumerate}[label=\textbf{Step \arabic*:}, ref=\textbf{Step \arabic*}, leftmargin=*, align=left]
\item The DSL compiler creates a prompt for the LLM based on the prompt template given to the \code{define} call.
\label{step:c1}
\item The DSL compiler sends this prompt to the LLM and receives the response from the LLM.
\label{step:c2}
\item The DSL compiler parses this response to extract the task's code and validates it.
\label{step:c3}
\end{enumerate}
\ref{step:c2} and \ref{step:c3} are executed multiple times until a generated code passes the validation in \ref{step:c3}.
The validation includes a syntactic check and a semantic check using execution with test examples.

\begin{figure}[t]
\begin{minipage}\linewidth
\begin{lstlisting}[style=mystyle,xleftmargin=2mm]
Q: Implement the following function:
```typescript
export function func({x, y}: {x: number, y: number}): number {
  // add 'x' and 'y'
}
```
\end{lstlisting}
\begin{lstlisting}[style=mystyle,xleftmargin=2mm]
A: 
```typescript
export function func({x, y}: {x: number, y: number}): number {
  // add 'x' and 'y'
  return x + y;
}
```
\end{lstlisting}
\begin{lstlisting}[style=mystyle,xleftmargin=2mm]
Q: Implement the following function:
```typescript
export function calculateFactorial({n}: {n: number}): number {
    // Calculate the factorial of 'n'
}
```
\end{lstlisting}
\end{minipage}
\caption{Prompt for asking the LLM to code the task} \label{fig:prompt}
\end{figure}

In \ref{step:c1}, the DSL compiler formulates a prompt to request the LLM to implement the task. This prompt instructs the LLM to complete the body of the function. The function signature is derived from the type information of the type parameter from the \code{define} call. Both the return type and parameter types are obtained from the \code{define} call's type parameter. The DSL compiler assigns a unique name to the function and outlines the empty function body for the LLM to fill in. We adopt a one-shot learning approach for function generation. In the generated prompt, we first provide a sample code generation to elucidate the code generation process. Then, we direct the LLM to generate a function implementing the specified task.

For instance, consider the scenario where the DSL compiler creates a function with the \code{define} API:

\noindent
\begin{minipage}\linewidth
\begin{lstlisting}[language=TypeScript, style=mystyle]
let calculateFactorial = define<number, {n: number}>("Calculate the factorial of {{n}}")
\end{lstlisting}
\end{minipage}
\noindent
The first and second type parameters specify the return type and parameter type of the defined function.
From this call, the DSL compiler generates a function whose signature is as follows:

\noindent
\begin{minipage}\linewidth
\begin{lstlisting}[language=TypeScript, style=mystyle]
function calculateFactorial({n}: {n: number}): number
\end{lstlisting}
\end{minipage}
We can call this function with a named argument, like \code{calculateFactorial(\{n: 10\})}.
We adopt named parameters instead of positional parameters since they are more robust for the modification of the prompt.
Named parameters are not affected by the appearance order in a template prompt.

The return type and parameter types originate from the type parameter of the \code{define} call. The DSL compiler assigns a unique name to the function and delineates the empty function for the LLM to complete.
The prompt that instructs the LLM to implement the function body is displayed in Figure \ref{fig:prompt}.
This prompt comprises three segments. The initial two segments are always the same regardless of the task.
They provide the LLM with an example of an input and output. This example entails constructing a function that accepts two numbers and outputs their sum.
The initial segment requests the LLM to implement the function.
While the function body is empty, the prompt details the task to be done as a comment inside the body.
The second segment gives an example of the response expected from the LLM.
The expected response is code that implements the function.
The third segment is the task-specific part and instructs the LLM to implement the given task.
This expected response is code that implements the function.
The structure of the instruction to the LLM is the same as the instruction in the first segment.

In \ref{step:c2}, the DSL compiler sends the created prompt to the LLM and receives the response from the LLM.
This step is executed using a low-level API provided by the LLM.
In our implementation, we use the OpenAI API for this step.
The API has a parameter named \code{temperature}, which controls the randomness of the response and can have a value ranging from 0.0 to 2.0.
We use the default value of 1.0 for this parameter, as we seek a certain level of randomness in the responses to ensure a unique response for each retry.

In \ref{step:c3}, the DSL compiler extracts the code from the response.
The LLM's reply is expected to contain the generated function in markdown's code block format: \code{```typescript ... ```}. As such, our DSL compiler can extract the function by finding the code block.
The DSL compiler checks the code syntactically and, optionally, checks it semantically by using the test examples provided in the \code{define} call.

The user can specify test examples for the task in the \code{define} call. These examples are provided as input-output pairs.
The input is a collection of key-value pairs, where the key is a parameter name and the value is a constant expression.
The output is a standalone constant expression.
The DSL compiler executes the generated function with the input and compares the output with the expected output.

If the code passes validation, the DSL compiler stores it in a file within the directory named \code{askit}, which is located in the same directory as the source file.
The DSL compiler also replaces the call to \code{define} with a call to the generated function.
The file containing the generated code is named after the template prompt.
The user can review the generated code if necessary.

\subsection{Interaction with an LLM for Directly Answerable Tasks} \label{sec:direct}

For each non-codable \code{define} and \code{ask} call, the DSL compiler just extracts the type information from the type parameter and encodes them into data to be used to generate a prompt at runtime.
Calls to functions defined by \code{define} and calls to \code{ask} are replaced with a call to our DSL runtime that takes the type information as a parameter in addition to the original parameters.
Our DSL runtime interacts with an LLM to execute the task specified by the prompt template passed to the \code{define} or \code{ask} call.
The steps of interaction between the DSL runtime and the LLM are as follows:
\begin{enumerate}[label=\textbf{Step \arabic*:}, ref=\textbf{Step \arabic*}, leftmargin=*, align=left]
\item The DSL runtime creates a prompt for the LLM based on the prompt template given to the \code{define} call.
\label{step:a1}
\item The DSL runtime sends this prompt to the LLM and receives the response from the LLM.
\label{step:a2}
\item The DSL runtime parses this response to extract the answer and validates it using the type information.
\label{step:a3}
\end{enumerate}
\ref{step:a2} and \ref{step:a3} are repeated until an answer in the valid type is available.

\begin{lstlisting}[style=mystyle, caption={Prompt to ask the LLM to perform the task directly},float=t,xleftmargin=2mm,label={lst:prompt_direct}]
You are a helpful assistant that generates responses in JSON format enclosed with ```json and ``` like:
```json
{ "reason": "Step-by-step reason for the answer", "answer": "Final answer or result" }
```
The response in the JSON code block should match the type defined as follows:
```ts
{ reason: string; answer: { title: string; author: string; year: number }[] }
```
Explain your answer step-by-step in the 'reason' field.

List 'n' classic books on 'subject'.
where 'n' = 5, 'subject' = "computer science"
\end{lstlisting}

In \ref{step:a1}, the AskIt runtime generates a prompt to ask the LLM to perform the task specified by the prompt template passed to the \code{define} or \code{ask} call.
It also uses the return type specified in the type parameter of the \code{define} or \code{ask} call to generate a prompt that constrains the LLM's response.

The difficulty of interacting with the LLM lies in extracting the answer from the LLM's response.
LLM responses are typically in natural language, making answer extraction challenging. To address this issue, we constrain the LLM's response to be in JSON (JavaScript Object Notation) format. The core idea of our prompt generation is to leverage the LLM's understanding of the grammar and semantics of programming languages.
For instance, an LLM, like GPT, can comprehend the grammar of JSON.
By requesting the LLM to answer in JSON format, we simplify the task of extracting the answer from its response.

However, merely specifying the JSON format does not guarantee the ease of answer extraction since the JSON structure may vary. This issue can be resolved by constraining the LLM's response to a specific JSON format. Fortunately, LLMs can grasp the semantics of types in programming languages. For instance, GPT can understand the semantics of TypeScript types. Furthermore, TypeScript types are ideal for constraining the JSON structure as they can be viewed as a JSON schema. For instance, the type \code{\{x: number; y: number\}} can be perceived as a JSON schema.
For example it accepts JSON object \code{\{"x": 1, "y": -1\}} but denies JSON object \code{[1, -1]}.
This approach is retained even when the host language is not TypeScript. Our AskIt implementation for Python uses TypeScript types to constrain the LLM's JSON response, even though Python is the host language.

We expect all JSON objects to have two fields: \code{reason} and \code{answer} regardless of the task.
Another possible option is not to use these fields and to use the entire JSON object as the answer.
However, if the \code{answer} field is not specified, the LLM sometimes automatically generates the \code{answer} field, which is not expected. This behavior complicates the answer extraction process. To avoid this issue, we always expect the LLM to provide the \code{answer} field.

As an example, consider a scenario where the LLM's response is expected to be a list of dictionaries. A function might be defined as follows:
\begin{lstlisting}[language=TypeScript, style=mystyle]
type Book = { title: string; author: string; year: number }
let getBooks = define<Book[]>("List {{n}} classic books on {{subject}}.")
\end{lstlisting}
Here, Line 1 defines a type \code{Book} and Line 2 defines a function \code{getBooks} that returns a list of \code{Book}. This function can be invoked as:
\begin{lstlisting}[language=TypeScript, style=mystyle]
let csBooks = getBooks({n: 5, subject: "computer science"})
\end{lstlisting}
When this function is called during runtime, the DSL runtime creates a prompt as displayed in Listing~\ref{lst:prompt_direct}.

In Listing~\ref{lst:prompt_direct}, the initial line indicates that the response should be in JSON format enclosed with \code{```json} and \code{```}. Lines 2–4 provide an example of the expected JSON format. We illustrate that the response should contain both an answer and a reason, exemplified by the provided response. Lines 1–4 are a fixed statement, always generated regardless of the function's parameters. Lines 5–8 are produced based on the function's type information. The 'reason' is always designated as \code{string}, regardless of the function's type information. Conversely, the 'answer' is task-specific. In this instance, the type of 'answer' is specified as \code{\{ title: string; author: string; year: number \}[]} since the function's type information is \code{Book[]}.

Line 9 is another fixed statement, always generated irrespective of the task description. We instruct the LLM to elucidate its answer in the 'reason' field. This promotes the Chain of Thought (CoT)~\cite{NEURIPS2022_9d560961}.

Lines 11-12 are constructed based on the prompt template passed to the \code{define} call and arguments passed to the function. \{\{ and \}\} in the prompt template are replaced with single quotes (Line~11), and the values of each parameter are appended to the prompt template (Line~12).

In \ref{step:a2}, the DSL runtime sends the prompt to the LLM and receives the response from the LLM.
This step uses the low-level API provided by the LLM. We use OpenAI API in our implmentation.

In \ref{step:a3}, the DSL runtime parses the response and extracts the answer from the response.
The response is expected to contain the JSON object.
However, the LLM does not always provide answers in the expected format.
In such cases, the DSL runtime requests the LLM to retry the task.
The DSL runtime employs a feedback mechanism to elicit the desired response in second or subsequent attempts.
Through multiple iterations, the DSL runtime refines the prompt to guide the LLM to provide the response in the expected format.
For each iteration, the DSL runtime evaluates the response based on the following criteria:
\begin{enumerate}
\item The response contains the JSON object.
\item The JSON object includes the \code{answer} field.
\item The \code{answer} field matches the expected type.
\end{enumerate}
If any of these criteria are not met, the DSL runtime refines the prompt by adding the LLM's response and a new instruction to the original prompt.
The new instruction points out the part of the response that does not meet the criteria and instructs the LLM to modify the response.

\subsection{Implementation for a Dynamically Typed Language} \label{sec:dynamic}

Our DSL compiler can be implemented in a dynamically typed language.
In a dynamically typed language, type information is provided at runtime.
Hence, the code generation for codable tasks should be done at runtime instead of at compilation time.

\begin{table*}[t]
\scriptsize
\centering
\caption{Types and their examples} \label{tab:types}
\begin{tabular}{llll}
\toprule
\code{API} & \code{Description} & \code{Usage Example} & \code{Equivalent Type in TypeScript} \\
\midrule
\code{int} & Integer & \code{int} & \code{number} \\
\code{float} & Floating Point Number & \code{float} & \code{number} \\
\code{bool} & Boolean & \code{bool} & \code{boolean} \\
\code{str} & String & \code{str} & \code{string} \\
\code{literal} & Literal & \code{literal(123)} & \code{123} \\
\code{list} & List & \code{list(int)} & \code{number[]} \\
\code{dict} & Dictionary & \code{dict(\{ 'x':int, 'y':int \})} & \code{\{x: number, y: number\}} \\
\code{union} & Union & \code{union(literal('yes'),literal('no'))} & \code{'yes' | 'no'} \\
\bottomrule
\end{tabular}
\end{table*}

Our implementation of AskIt for Python is fully realized as a library.
The API of AskIt for Python is almost identical to the API of AskIt for TypeScript, except for the following two points:
\begin{enumerate}
\item The return type of \code{ask} and \code{define} is specified as a parameter of the function rather than as a type parameter.
\item Compilation is invoked explicitly by calling the \code{compile} method on the function returned by \code{define}.
\end{enumerate}
The first point concerns how the type information is provided to \code{define}.
In the case of the Python implementation, the type is specified by a type object provided as the first argument of the function rather than a type parameter.
AskIt for Python offers APIs to create a type object for the return type of \code{ask} and \code{define}.
The provided APIs are listed in Table~\ref{tab:types}.
The first column is the name of the API, and the second column describes the type created by the API.
The third column provides usage examples, and the fourth column indicates the equivalent type in TypeScript.

For instance, the same task introduced in \ref{sec:direct} can be implemented in Python as follows:
\begin{lstlisting}[language=Python, style=mystyle, escapechar=@]
Book = dict({ "title": str, "author": str, "year": int })
getBooks = define(List(Book), "@\color{codepurple}List@ {{n}} classic books on {{subject}}.")
\end{lstlisting}
The first line defines a type object using the provided APIs.
The second line defines a function that returns a list of \code{Book}.

The second point of difference concerns how code generation is conducted.
In Python's case, users must explicitly specify when code generation occurs.
For this purpose, functions defined by \code{define} return a function object that implements the \code{compile} method.
When the \code{compile} method is invoked, code generation proceeds in the same manner as the compilation time of AskIt for TypeScript.

For instance, the task described in Section \ref{sec:code_generation} can be implemented in Python as follows:
\begin{lstlisting}[language=Python, style=mystyle]
calculateFactorial = define(int, "Calculate the factorial of {{n}}").compile()
\end{lstlisting}
When the \code{compile} method is invoked, code generation takes place, resulting in the return of a function object that implements the task. The generated code is cached in a file upon its initial creation, ensuring that code generation happens only once, regardless of how many times the \code{compile} method is called.

\section{Experimental Evaluation} \label{sec:evaluation}
To evaluate the effectiveness of AskIt, we conducted a series of experiments. Each experiment was designed to answer distinct questions about our DSL, specifically targeting different task types:

\begin{itemize}
\item Codable tasks:
\begin{itemize}
\item \textbf{RQ1:} How does AskIt reduce the LOC required to implement codable tasks?
\item \textbf{RQ2:} Are examples of tasks effective for improving the accuracy of generated code?
\end{itemize}

\item Directly answerable tasks:
\begin{itemize}
\item \textbf{RQ3:} How does AskIt reduce the LOC of prompt generation for directly answerable tasks?
\end{itemize}

\item Intersecting tasks:
\begin{itemize}
\item \textbf{RQ4:} How does the speed and performance of functions generated by AskIt compare to the same function before code generation?
\end{itemize}
\end{itemize}
\noindent
To address these questions, we carried out three different experiments for each task category.




\subsection{Codable Tasks}
\subsubsection{Common Coding Tasks} \label{sec:common_tasks}
To address RQ1 and RQ2, we designed an experiment that involved implementing a set of 50 tasks using AskIt. To ensure these tasks were both relevant and realistic, we enlisted the help of ChatGPT.
Specifically, we inquired about the 50 most commonly requested TypeScript coding tasks.
These 50 tasks subsequently served as the foundation for our implementation in TypeScript and Python using AskIt. To verify the correctness of the generated code, we supplied AskIt with example tests for each task. If a test failed, AskIt would attempt code regeneration up to a predefined maximum retry limit, which was set to 9.
In this experiment, we specified "gpt-3.5-turbo-16k" as the backend LLM for AskIt.

Our results for the first ten tasks and additional remarkable tasks are presented in table \ref{tab:common_tasks}.
The first column enumerates the 50 tasks.
The table's second column displays the template prompt used in both TypeScript and Python implementations.
The third column indicates the return type utilized in the \code{define} call for each task.
The fourth column delineates the parameter types utilized in the \code{define} call for each task.
We only use parameter types for TypeScript since Python implementation does not use parameter types.
Columns five and six enumerate the lines of code (LOC) in the generated TypeScript code and the associated retries.
LOC counts only substantive lines, omitting empty lines or comment-only lines.
The next two columns present analogous details for Python.

On average, AskIt produced 7.56 lines for TypeScript and 6.52 lines for Python. 
Considering that each AskIt function definition resulted in a single line, an effective reduction of 6.56 and 5.52 lines was achieved for TypeScript and Python, respectively.
Although all tasks were successfully rendered in TypeScript, tasks \#11 and \#21-24 encountered issues in Python. This stems from the Python variant of AskIt not leveraging parameter types for prompt generation in the LLM. For instance, in Task \#11 for Python, we presumed the parameter type for xs was \code{Array}. Contrarily, the generated code assumed it was \code{set}.

The retry count for successful code generation varied between 0 and 7 for Python. In a few instances, the initially generated code did not pass the example test. Even though the retry count seems negligible, it's imperative to recognize that it's not consistently zero. This indicates that the LLM can occasionally produce erroneous code. As an example, the code for Task \#14 in Python failed its initial run, computing the Fibonacci numbers up to n + 1 rather than n, necessitating seven retries. Thus, supplying AskIt with task examples is vital for assuring the correctness of the generated code.
\begin{table*}[t]
  \centering
  \tiny
  \caption{Summary of the 50 codable tasks implemented using AskIt} \label{tab:common_tasks}
  \resizebox{\textwidth}{!}{%
  \tiny
  \begin{tabular}{clllllll} 
  \toprule
  \#
  &
  Template Prompt
  &
  Return Type in TypeScript
  &
  Parameters Types (TypeScript Only)
  &
  \multicolumn{2}{c}{TypeScript}
&
\multicolumn{2}{c}{Python}
\\
&
&
&
&
LOC & Retry
&
LOC & Retry
  \\ \midrule
  \input{tab/common_tasks.tex}
\\  \bottomrule
  \end{tabular}
  }
  \end{table*}

\subsubsection{HumanEval} \label{sec:humaneval}

To compare the code generated by AskIt with the hand-written code, we used the HumanEval benchmark \cite{chen2021evaluating}.
The HumanEval benchmark is a dataset of 164 coding tasks.
Each task description includes a prompt, test cases, and its corresponding hand-written solution.
The test cases are input-output pairs, and the hand-written code is a Python function that passes all the test cases. The HumanEval benchmark is designed to evaluate the performance of LLMs on coding tasks. We used the HumanEval benchmark to compare the code generated by AskIt with the hand-written code.

We specified the prompts of the HumanEval benchmark as the prompt templates for AskIt.
We converted the few-shot learning examples described in the prompts into training examples for the \code{define} call.
We used the test cases of the HumanEval benchmark as test examples to check the correctness of the generated code.

As a result, we successfully generated valid code for 139 out of 164 tasks, which means the success rate is 84.8~\%.
This accuracy is comparable to the accuracy of state-of-the-art LLMs~\cite{hong2023metagpt,zelikman2023parsel,huang2023anpl,muennighoff2023octopack,zhou2023language} on the HumanEval benchmark.
On average, the generated code for the 139 tasks was 8.05 lines, while the source code in AskIt was 23.74 lines.
The source code is longer than the generated code because it includes additional lines for training and test examples.
Compared to the hand-written code, the generated code was on average 1.27 times longer, as the average length of the hand-written code was 7.57 lines.

\begin{figure}
\centering
\includegraphics[width=0.30\textwidth]{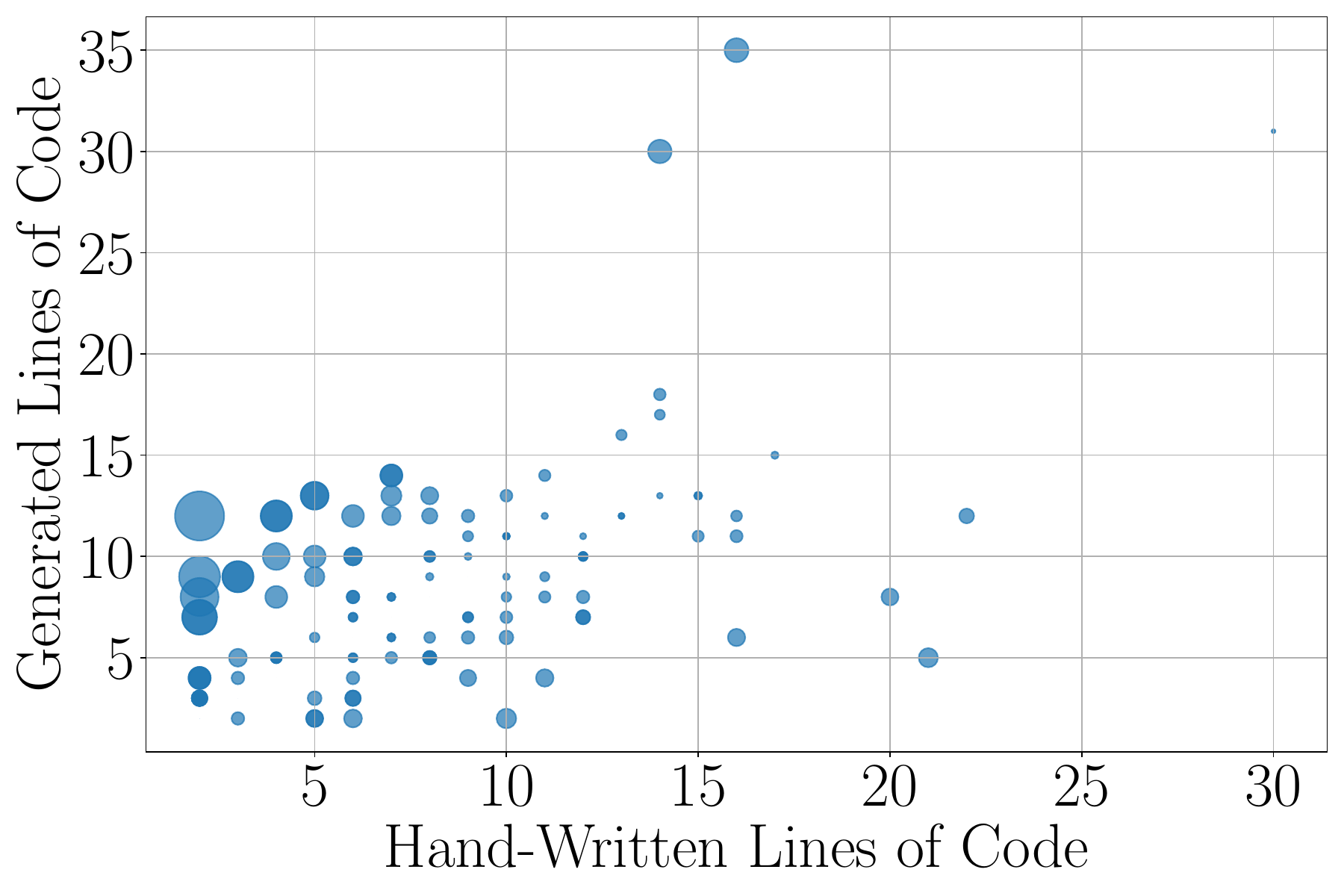}
\caption{Scatter plot of LOC of generated code and hand-written code} \label{fig:loc}
\end{figure}

Figure \ref{fig:loc} shows the scatter plot of the LOC (Lines of Code) of the generated code and the hand-written code.
Each point on the plot corresponds to a specific coding task, with the x-axis denoting the LOC in the hand-written version and the y-axis representing the LOC in the generated version. The size of each point is proportional to the ratio of the generated LOC to the hand-written LOC. This scatter plot reveals a broad distribution of data points, indicating variability in the relationship between the lengths of the hand-written and generated code.
Although AskIt generally produces longer code than the hand-written code, in 49 out of 139 tasks (35.3~\%) the generated code is shorter.
This result indicates that AskIt can generate shorter code than the hand-written code in some cases.


\subsection{Directly Answerable Tasks} \label{sec:openai_evals}

To answer RQ3, we transformed existing prompts for LLMs into AskIt prompts tailored for directly answerable tasks. We then compared the lengths of the original prompts with those of the AskIt prompts. Our source of these prompts was the OpenAI Evals\footnote{\url{https://github.com/openai/evals}}. The OpenAI Evals repository contains over 300 benchmarks, representing real-world use cases of LLMs. Notably, a majority of these benchmarks originate from real-world LLM users.

Each benchmark in the repository consists of multiple test cases.
In turn, each test case includes a prompt and the anticipated LLM response. For this experiment, we restricted our focus to the first 50 benchmarks from OpenAI Evals. Additionally, we selected only the first test case from each benchmark, given that all the test cases within a particular benchmark share a similar type but with varying inputs.

Our modification process for the AskIt prompts involved eliminating superfluous information. This includes phrases dictating the LLM's response format or prompting the LLM to elucidate its answer's rationale. The AskIt prompt inherently incorporates such information. For instance, consider this excerpt from the original prompt for the benchmark \emph{2d\_movement.dev.v0}: \code{"Please note: In the following EXERCISE, it is essential that you only respond with a single line in the format (x, y)."}
Such directives can be omitted from the AskIt prompt. Instead, we detail the expected LLM response type within the AskIt prompt. In the aforementioned case, the response type is designated as \code{\{ x: number, y: number \}}.

Given that most benchmarks were unsolvable by GPT-3.5 and GPT-4, we solely ensured that our modified prompt yielded an output format congruent with the LLM's expected response, as specified in the test case.

\begin{figure}
\centering
\includegraphics[width=0.48\textwidth]{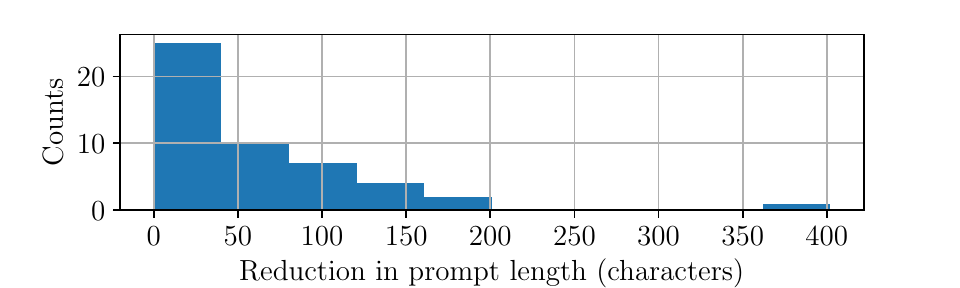}
\caption{Histogram of character count reductions in AskIt versus original prompts} \label{fig:prompt_reductions}
\end{figure}

Figure \ref{fig:prompt_reductions} presents a histogram summarizing the reductions in prompt lengths. The x-axis represents the range of character counts, and the y-axis displays the frequency of benchmarks within each range. This histogram visually illustrates the distribution of prompt length reductions achieved by transforming the original prompts into AskIt prompts. On average, we observed a 16.14~\% reduction in character count from the original prompts.

\begin{figure}
\centering
\includegraphics[width=0.48\textwidth]{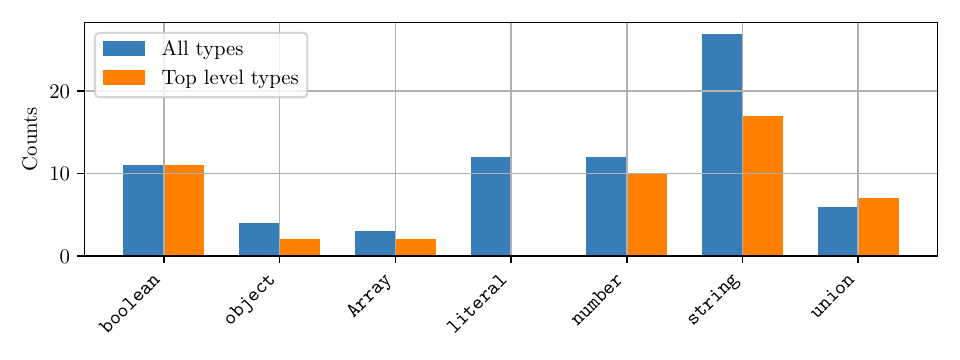}
\caption{Number of uses for each type} \label{fig:type_count}
\end{figure}

All the types used in the benchmarks are presented in Figure \ref{fig:type_count}.
The x-axis represents the types, and the y-axis displays the number of uses for each type.
We count the number of uses in two ways: (1) counting the number of uses of the top-level type, and (2) counting the number of uses of all types.
The most frequently used top-level type is \code{string}, followed by \code{number} and \code{boolean}.
Although the literal type is not a top-level type, it is frequently combined with other types

\subsection{Intersecting Tasks} \label{sec:gsm8k}

One of the benefits of the unified interface provided by AskIt is that it improves the performance of intersecting tasks by using the LLM to generate code for the task without rewriting the prompt template.
To answer RQ4, we compared the performance of functions in two scenarios: when using LLMs to answer the task directly, and when the function is generated by the LLM to perform the task.

\begin{table}[t]
\centering
\footnotesize
\caption{Experimental results using GSM8K}\label{tab:gsm8k}  
\input{tab/merged_tab.tex}  
\end{table}

We used the GSM8K \cite{DBLP:journals/corr/abs-2110-14168} benchmark, a dataset of high-quality grade school math word problems.
We converted numerical values surrounded by spaces in the problem description into variables since the generated programs are often reused with different values in a practical setting.

We used the original values as test examples to check the correctness of the generated program.
We execute the program in the following two ways: (1) directly executing the program generated by AskIt, and (2) compiling the program generated by AskIt and executing the compiled program.
If the generated program failed to pass the test example, we retried up to 9 times to generate the correct program.

The GSM8K benchmark consists of training data and test data.
We only use the test data for our evaluation since we use GPT without fine-tuning with the training data.
The test data contained 1,319 problems.
We used "gpt-4" as the backend LLM for AskIt in this experiment.
All the time measurements were conducted on a machine with an Apple M1 CPU and 16GB of RAM.

In TypeScript, \placeholder{1,138} problems were solved by GPT-4.
In Python, GPT-4 directly solved \placeholder{1,159} problems.
The difference is not significant.
Since both implementations use the same GPT model and the same prompt, the difference seems to come from the randomness of the response of GPT-4.
We use these \placeholder{1,138} and \placeholder{1,159} problems for program generation.
We successfully generated the program for \placeholder{1,114} and \placeholder{1,134} problems in TypeScript and Python, respectively.

The results are shown in Table~\ref{tab:gsm8k} for TypeScript and Python, respectively.
The latency is the time to obtain the answer from the LLM.
The execution time is the time to execute the generated function.
The compilation time is the time to generate the function for the problem.
On average, the generated codes answered the problem 275,092.55x and 6,969,904.73x times faster in TypeScript and Python, respectively, than using the LLM directly to answer the problem.
While the speedup ratio is different between TypeScript and Python, the generated code is significantly faster than the LLM in both cases.


\section{Related Work} \label{sec:related_work}



LMQL~\cite{10.1145/3591300} is a query language specifically designed for large language models (LLMs), combining natural language prompts with the expressiveness of Python. It provides features such as constraints, debugging, retrieval, and control flow to facilitate interaction with LLMs. LMQL offers full Python support, enabling powerful control flow and logic in a prompting logic. LMQL allows model developers to declare logical constraints governing model output. These get turned into "token-level prediction masks" - tokens being what LLMs deal with. While it supports type constraints, the supported types are limited and not integrated with the type system of the underlying programming language. For example, LMQL does not support the ability to define custom types.
LLMChain\footnote{\url{https://docs.langchain.com/docs/components/chains/llm-chain}} is a library that provides a prompt template that supports parameters like AskIt.
It generates prompts by filling in the template with the parameters.
While LLMChain provides \code{apply_and_parse} function to parse the response of LLM, the user needs to specify the parser to extract the answer from the response.
On the other hand, AskIt automatically parses the response and extracts the answer from the response based on the type information.
The OpenAI API\footnote{\url{https://beta.openai.com/docs/api-reference}} is a low-level API for LLMs.
The latest version of the OpenAI API supports function calling.
The user can register a user-defined function with the LLM and call it from the prompt.
The argument for the function is passed as a JSON object.
By exploiting this feature, users can obtain answers in a specific format.
However, the code is more complicated than using AskIt, as the user needs to write code to register the function, and make it callable from the prompt.
While AskIt does not use this feature, it can be used to implement AskIt.

Unlike AskIt, LMQL, LLMChain, and the OpenAI API do not support code generation.
While code can be generated using them if the user writes the prompt accordingly, the generated code cannot be seamlessly integrated into the rest of the program.
With AskIt, switching between using the LLM directly and using the generated code can be done without changing the prompt template.

Another approach to integrating LLMs into programming is to enable LLMs to use APIs so that they can access broader and more dynamic knowledge bases, as well as perform complex computational tasks. The challenge is the complexity of integrating millions of changing APIs, which can have overlapping functionalities and nuanced limitations. Gorilla~\cite{patil2023gorilla} proposes using self-instruct fine-tuning and retrieval to enable LLMs to accurately select from large, overlapping, and changing sets of tools expressed via their APIs and API documentation.

\section{Future Work}

One limitation of AskIt is that it does not guarantee the safety of the generated code.
If the generated function contains harmful code, it can pose a security risk.
For instance, the generated function might unexpectedly contain code that deletes all files in a directory.
The current implementation relies on the user's review of the generated code.
To mitigate this risk, it is necessary to develop a mechanism that ensures the safety of the generated code.
Possible approaches include using a sandbox or a static analysis tool.

Another improvement would be to generate more efficient code.
Although we currently generate code in TypeScript and Python, this code is not optimized.
One potential approach involves utilizing LLMs to generate code in low-level languages, such as C or LLVM IR, with a foreign function interface.

\section{Conclusion} \label{sec:conclusion}
In this paper, we introduced a domain-specific language (DSL), \emph{AskIt}.
AskIt provides a unified interface for interacting with large language models (LLMs) for various tasks.
The unified interface supports (1) Type-guided output control of LLMs, (2) Template-based function definition, and (3) Code generation for codable tasks. 
Experimental results show that AskIt reduces the number of lines of code required to implement codable tasks by 6.56 and 5.52 lines for TypeScript and Python, respectively. AskIt also reduces the number of lines of prompt generation for directly answerable tasks by 16.14~\%. The generated code for intersecting tasks is 275,092.55x and 6,969,904.73x times faster on average than using the LLM directly to answer the problem in TypeScript and Python, respectively.

\input{ae}

\bibliographystyle{IEEEtran}
\balance
\bibliography{references} 

\end{document}

%% file: tab/common_tasks.tex
1 &
    \begin{lstlisting}
    Reverse the string {{s}}.
    \end{lstlisting}
    &
    \begin{lstlisting}
    string
    \end{lstlisting}
    &
    \begin{lstlisting}
    { s: string }
    \end{lstlisting}
    & 5
    & 0
    & 4
    & 0
    \\ 
2 &
    \begin{lstlisting}
    Calculate the factorial of {{n}}.
    \end{lstlisting}
    &
    \begin{lstlisting}
    number
    \end{lstlisting}
    &
    \begin{lstlisting}
    { n: number }
    \end{lstlisting}
    & 9
    & 0
    & 7
    & 0
    \\ 
3 &
    \begin{lstlisting}
    Concatenate the strings {{ss}}.
    \end{lstlisting}
    &
    \begin{lstlisting}
    string
    \end{lstlisting}
    &
    \begin{lstlisting}
    { ss: string[] }
    \end{lstlisting}
    & 5
    & 0
    & 5
    & 0
    \\ 
4 &
    \begin{lstlisting}
    Sort the numbers {{ns}} in ascending order.
    \end{lstlisting}
    &
    \begin{lstlisting}
    number[]
    \end{lstlisting}
    &
    \begin{lstlisting}
    { ns: number[] }
    \end{lstlisting}
    & 5
    & 0
    & 5
    & 0
    \\ 
5 &
    \begin{lstlisting}
    Find the largest number in {{ns}}.
    \end{lstlisting}
    &
    \begin{lstlisting}
    number
    \end{lstlisting}
    &
    \begin{lstlisting}
    { ns: number[] }
    \end{lstlisting}
    & 3
    & 0
    & 5
    & 0
    \\ 
6 &
    \begin{lstlisting}
    Check if {{n}} is a palindrome.
    \end{lstlisting}
    &
    \begin{lstlisting}
    boolean
    \end{lstlisting}
    &
    \begin{lstlisting}
    { n: number }
    \end{lstlisting}
    & 7
    & 0
    & 6
    & 0
    \\ 
7 &
    \begin{lstlisting}
    Calculate the sum of all numbers in {{ns}}.
    \end{lstlisting}
    &
    \begin{lstlisting}
    number
    \end{lstlisting}
    &
    \begin{lstlisting}
    { ns: number[] }
    \end{lstlisting}
    & 5
    & 0
    & 5
    & 0
    \\ 
8 &
    \begin{lstlisting}
    Calculate the average of all numbers in {{ns}}.
    \end{lstlisting}
    &
    \begin{lstlisting}
    number
    \end{lstlisting}
    &
    \begin{lstlisting}
    { ns: number[] }
    \end{lstlisting}
    & 8
    & 0
    & 5
    & 0
    \\ 
9 &
    \begin{lstlisting}
    Count the number of occurrences of {{x}} in {{xs}}.
    \end{lstlisting}
    &
    \begin{lstlisting}
    number
    \end{lstlisting}
    &
    \begin{lstlisting}
    { xs: number[]; x: number }
    \end{lstlisting}
    & 6
    & 0
    & 8
    & 0
    \\ 
10 &
    \begin{lstlisting}
    Remove all instances of {{x}} from {{xs}}.
    \end{lstlisting}
    &
    \begin{lstlisting}
    number[]
    \end{lstlisting}
    &
    \begin{lstlisting}
    { xs: number[]; x: number }
    \end{lstlisting}
    & 3
    & 0
    & 6
    & 0
    \\ 
11 &
    \begin{lstlisting}
    Return the unique elements in {{xs}}.
    \end{lstlisting}
    &
    \begin{lstlisting}
    number[]
    \end{lstlisting}
    &
    \begin{lstlisting}
    { xs: number[] }
    \end{lstlisting}
    & 5
    & 0
    & 0
    & 0
    \\ 
12 &
    \begin{lstlisting}
    Find the factorial of {{n}}.
    \end{lstlisting}
    &
    \begin{lstlisting}
    number
    \end{lstlisting}
    &
    \begin{lstlisting}
    { n: number }
    \end{lstlisting}
    & 9
    & 0
    & 7
    & 0
    \\ 
14 &
    \begin{lstlisting}
    Generate the Fibonacci sequence up to {{n}}.
    \end{lstlisting}
    &
    \begin{lstlisting}
    number[]
    \end{lstlisting}
    &
    \begin{lstlisting}
    { n: number }
    \end{lstlisting}
    & 10
    & 0
    & 19
    & 0
    \\ 
21 &
    \begin{lstlisting}
    Convert the JSON object {{o}} into a string.
    \end{lstlisting}
    &
    \begin{lstlisting}
    string
    \end{lstlisting}
    &
    \begin{lstlisting}
    { o: any }
    \end{lstlisting}
    & 3
    & 0
    & 0
    & 0
    \\ 
24 &
    \begin{lstlisting}
    Find the difference between the dates {{d1}} and {{d2}}.
    \end{lstlisting}
    &
    \begin{lstlisting}
    number
    \end{lstlisting}
    &
    \begin{lstlisting}
    { d1: Date; d2: Date }
    \end{lstlisting}
    & 6
    & 0
    & 0
    & 0
    

%% file: tab/merged_tab.tex
\begin{tabular}{lrr}
\toprule
Average Metrics & TypeScript & Python \\
\midrule
Latency (s) & 13.28 & 22.97 \\
Execution Time (\si{\micro\second}) & 49.11 & 5.09 \\
Compilation Time (s) & 14.19 & 20.38 \\
Speedup Ratio & 275092.55 & 6969904.73 \\
\bottomrule
\end{tabular}

%% file: ae.tex
\appendix

\lstset{
  basicstyle=\small\ttfamily, 
}

\section{Artifact Appendix}

\subsection{Abstract}

The appendix is designed to facilitate the replication of experiments conducted in this paper. It includes a Docker container, complemented by a set of scripts, to seamlessly replicate the results shown in Figure~\ref{fig:loc}, Figure~\ref{fig:prompt_reductions}, and Figure~\ref{fig:type_count}, as well as the data presented in Table~\ref{tab:common_tasks} and Table~\ref{tab:gsm8k}.

\subsection{Artifact check-list (meta-information)}

{\small
\begin{itemize}
  \item {\bf Data set: HumanEval, OpenAI Evals, GSM8K}
  \item {\bf Run-time environment: Any operating system that supports Docker}
  \item {\bf Hardware: Any machine with Docker installed}
  \item {\bf Metrics: Lines of code, reduction in prompt length, speedup}
  \item {\bf Output: Figure~\ref{fig:loc}, \ref{fig:prompt_reductions}, \ref{fig:type_count}, Table~\ref{tab:common_tasks}, Table~\ref{tab:gsm8k}}
  \item {\bf Experiments: }
  \item {\bf How much disk space required (approximately)?: 5GB}
  \item {\bf How much time is needed to prepare workflow (approximately)?: 15 minutes}
  \item {\bf How much time is needed to complete experiments (approximately)?: 1 day}
  \item {\bf Publicly available?: Yes}
  \item {\bf Code licenses (if publicly available): MIT License}
  \item {\bf Workflow framework used?: No, but we provide scripts to reproduce the experiments}
  \item {\bf Archived (provide DOI)?: Yes}
\end{itemize}

\subsection{Description}

\subsubsection{How delivered}

This artifact is available at \url{https://github.com/katsumiok/askit-artifact.git} and archived at \url{https://doi.org/10.5281/zenodo.10327179}~\cite{myArtifact2023}.

\subsubsection{Hardware dependencies}

No hardware dependencies.

\subsubsection{Software dependencies}

The artifact is delivered as a Docker container. The only software dependency is Docker.

\subsubsection{Data sets}

The artifact includes the following data sets:
\begin{itemize}
\item HumanEval
\item OpenAI Evals
\item GSM8K
\end{itemize}
These data sets are modified to be compatible with AskIt.

\subsection{Installation}

\begin{enumerate}
\item Clone the artifact from the GitHub repository.
  
\begin{lstlisting}[language=bash]
$ git clone https://github.com/katsumiok/askit-artifact.git
\end{lstlisting}
\item Install Docker.
\item Edit \code{askit-artifact/Dockerfile} to specify a API key for OpenAI API.
Update the following line in \code{askit-artifact/Dockerfile}:
\begin{lstlisting}[language=bash]
ENV OPENAI_API_KEY sk-xxxxxxxxxxxxxxxxxxxxxxxx
\end{lstlisting}
with your API key.

\item Build a Docker image from the Dockerfile in the artifact directory.

\begin{lstlisting}[language=bash]
$ cd askit-artifact
$ docker build -t askit .
\end{lstlisting}
\end{enumerate}

\subsection{Experiment workflow}

Before running the experiments, you need to start the Docker container by running the following command in the artifact directory:
\begin{lstlisting}[language=bash]
$ docker run -it -v $PWD:/root/docker-artifact askit
\end{lstlisting}

To reproduce the results shown in Figure~\ref{fig:loc}, Figure~\ref{fig:prompt_reductions}, and Figure~\ref{fig:type_count}, as well as the data presented in Table~\ref{tab:common_tasks} and Table~\ref{tab:gsm8k}, run the following commands:
\begin{lstlisting}[language=bash]
$ make  
\end{lstlisting}

The above command runs scripts to generate the figures and tables from the experimental results already included in the artifact.

To reproduce the experimental results from scratch, remove the intermediate files by running the following command:
\begin{lstlisting}[language=bash]
$ make clean_all
\end{lstlisting}
and then run the following command:
\begin{lstlisting}[language=bash]
$ ./run_all.sh
\end{lstlisting}
The above command runs all the experiments and generates intermediate files used to generate the figures and tables.
Typing \code{make} again generates the figures and tables from the intermediate files.
Note that \code{make} does not run the experiments and only generates the figures and tables from the intermediate files.

\subsection{Evaluation and expected result}

The workflow generates the files listed in Table~\ref{tab:files}.
These figures and tables may be slightly different from the ones shown in the paper due to the randomness of the language models and the difference in the environment.
However, the figures and tables should be similar to the ones shown in the paper.

\begin{table}[t]
\scriptsize
\centering
\caption{Files Generated by the Experiment Workflow}
\label{tab:files}
\begin{tabular}{ll}
\toprule
File & Figure/Table \\
\midrule
\code{fig/loc.pdf} & Figure~\ref{fig:loc} \\
\code{fig/prompt\_reduction.pdf} & Figure~\ref{fig:prompt_reductions} \\
\code{fig/type\_count.pdf} & Figure~\ref{fig:type_count} \\
\code{tab/common\_tasks.tex} & Table~\ref{tab:common_tasks} \\
\code{tab/gsm8k.tex} & Table~\ref{tab:gsm8k} \\
\bottomrule
\end{tabular}
\end{table}

\subsection{Experiment customization}

\begin{table*}[t]
\centering

\caption{Scripts to Customize the Experiments}
\label{tab:scripts}
\begin{tabular}{llll}
\toprule
Script & Description & Section  \\
\midrule
\code{coding/ts-askit/examples/run.sh} & Run AskIt on 50 common coding tasks in TypeScript & Section~\ref{sec:common_tasks}\\
\code{coding/pyaskit/examples/run.sh} & Run AskIt on 50 common coding tasks in Python & Section~\ref{sec:common_tasks}\\
\code{HumanEval/run.sh} & Run AskIt on HumanEval in Python & Section~\ref{sec:humaneval}\\
\code{openai\_evals/run.sh} & Run AskIt on OpenAI Evals in Python & Section~\ref{sec:openai_evals}\\
\code{GSM8K/ts/run.sh} & Run AskIt on GSM8K in TypeScript & Section~\ref{sec:gsm8k}\\
\code{GSM8K/python/run.sh} & Run AskIt on GSM8K in Python & Section~\ref{sec:gsm8k}\\
\bottomrule
\end{tabular}
\end{table*}
  
Table~\ref{tab:scripts} shows the scripts to customize the experiments.
These scripts are executed by \code{run_all.sh}.
Each script sets the environment variable \code{ASKIT\_MODEL} to specify the language model used in the experiments.
By setting the environment variable \code{ASKIT\_MODEL} in these scripts, you can customize the experiments to use a different model.
For example, we specify "gpt-4" in \code{GSM8K/ts/run.sh} to use GPT-4 for the experiments on GSM8K in TypeScript.
You can change it to "gpt-3.5-turbo-16k" to use GPT-3.5 Turbo 16K instead.

The remaining part of this section describes the details of the experiments in each section.

\subsubsection{Experiment in \ref{sec:common_tasks}: 50 Common Coding Tasks}

In this experiment, we run AskIt on 50 common coding tasks in TypeScript and Python.
The source code of the tasks is included in the artifact in the following files:
\begin{itemize}
\item \code{coding/ts-askit/examples/src/top50def.ts}
\item \code{coding/pyaskit/examples/top50.py}
\end{itemize}

\code{coding/ts-askit/examples/run.sh} and \code{coding/pyaskit/examples/run.sh} run AskIt on the tasks in TypeScript and Python, respectively.
After running these scripts, the generated code is saved in \code{coding/ts-askit/examples/src/askit} and \code{coding/pyaskit/examples/askit}, respectively.
\code{coding/make_table.py} generates Table~\ref{tab:common_tasks} from the generated code.

\subsubsection{Experiment in \ref{sec:humaneval}: HumanEval}

In this experiment, we run AskIt on HumanEval in Python and compare the generated code with the hand-written code.
There are 164 tasks in HumanEval.
\code{HumanEval/HumanEval.jsonl} is the original HumanEval data, which contains the hand-written code and the prompt.
The 164 tasks written in AskIt are represented by files named 0.py to 163.py in the \code{HumanEval/HumanEval} directory.
After running \code{HumanEval/run.sh}, the generated code is saved in \code{HumanEval/askit}.
\code{HumanEval/make_table.py} generates Figure~\ref{fig:loc} from the generated code and the hand-written code.

\subsubsection{Experiment in \ref{sec:openai_evals}: OpenAI Evals}

In this experiment, we compare the lengths of the prompts generated by AskIt with those of the original prompts in OpenAI Evals. The prompts for AskIt are stored in \code{openai_evals/data}, and the original prompts are stored in \code{openai_evals/odata} in JSON format. \code{openai_evals/run.sh} runs AskIt on the prompts in \code{openai_evals/data} to check if the responses of the LLMs correspond to the expected types. \code{openai_evals/make_table.py} generates Figure~\ref{fig:prompt_reductions} and Figure~\ref{fig:type_count} by comparing the generated prompts with the original prompts.

\subsubsection{Experiment in \ref{sec:gsm8k}: GSM8K}

In this experiment, we run GSM8K in TypeScript and Python in two settings: with and without code generation. There are 1319 tasks in GSM8K. Files named from 0.ts to 1318.ts in the \code{GSM8K/ts/src} directory contain the 1319 tasks written in TypeScript and are used to run GSM8K without code generation. Similarly, files named from 0.ts to 1318.ts in the \code{GSM8K/ts/src2} directory are the tasks written in TypeScript, used to run GSM8K with code generation. \code{GSM8K/ts/run.sh} transpiles the TypeScript files in both \code{GSM8K/ts/src} and \code{GSM8K/ts/src2} and runs the generated JavaScript files. However, \code{GSM8K/ts/run.sh} only generates functions for the tasks in \code{GSM8K/ts/src2} and does not generate functions for the tasks in \code{GSM8K/ts/src}. After running \code{GSM8K/ts/run.sh}, the generated code is saved in \code{GSM8K/ts/src2/askit}. Metrics, including the execution times, are saved in \code{GSM8K/ts/json} and \code{GSM8K/ts/json2} as JSON files named from 0.json to 1318.json, respectively.

A file named \code{GSM8K/test.jsonl} contains task descriptions in JSON format. \code{GSM8K/python/run.py}, executed by \code{GSM8K/python/run.sh}, runs GSM8K in Python using \code{GSM8K/test.jsonl}. After running \code{GSM8K/python/run.py}, the generated code is saved in \code{GSM8K/python/askit}. Metrics, including execution times, are saved in \code{GSM8K/python/json} as JSON files named from 0.json to 1318.json. These JSON files contain the metrics for both the tasks with and without code generation.

\code{GSM8K/make_table.py} generates Table~\ref{tab:gsm8k} from the metrics produced by \code{GSM8K/ts/run.sh} and \code{GSM8K/python/run.sh}.